\documentclass[12pt]{article}
\usepackage{epsf}
\begin{document}
\def\@fpsep{5mm}
\begin{center}
{\bfseries DIRECT CP VIOLATION IN NONLEPTONIC KAON DECAYS\\
           BY AN EFFECTIVE CHIRAL LAGRANGIAN APPROACH AT $O(p^6)$%
\footnote{Talk at the XV Int. Seminar on High Energy Physics Problems
"Relativistic Nuclear Physics \& Quantum Chromodynamics",
Dubna, September 25--29, 2000.}
}

\vskip 5mm

A.A. Bel'kov$^{1 \dag}$, G. Bohm$^{2}$, A.V. Lanyov$^{1}$ and A.A. Moshkin$^{1}$

\vskip 5mm

{\small
(1) {\it Particle Physics Laboratory, Joint Institute for Nuclear Research,\\
                      141980 Dubna, Moscow region, Russia}
\\
(2) {\it DESY--Zeuthen, Platanenallee 6, D--15735 Zeuthen, Germany}
\\
$\dag$ {\it E-mail: belkov@cv.jinr.dubna.su}}
\end{center}

\vskip 5mm

\begin{center}
\begin{minipage}{150mm}
\centerline{\bf Abstract}
    A self-consistent analysis of $K\to 2\pi$ and $K\to 3\pi$ decays within 
a unique framework of chiral dynamics applied to the QCD-corrected weak 
nonleptonic quark lagrangian has been performed.
    The results on $K\to 2\pi$ amplitudes at $O(p^6)$, including the 
value for $\varepsilon^{'}/\varepsilon$, are compared with the
experiment to fix phenomenological B-factors for mesonic
matrix elements of nonpenguin and penguin four-quark operators. 
    The dependence of B-factors on  different theoretical uncertainties and
experimental errors of various input parameters is investigated.
    Finally, we present our estimates at $O(p^6)$ for the CP-asymmetry of 
linear slope parameters in the $K^{\pm}\to 3\pi$ Dalitz plot.
\\
{\bf Keywords:}
              direct CP violation, nonleptonic decays, K-mesons, chiral model
\end{minipage}
\end{center}

\vskip 10mm

    The starting point for most calculations of nonleptonic kaon decays 
is an effective weak lagrangian of the form
\cite{vzsh,gilman-wise}
\begin{equation}
{\cal L}^{q}_{w}\big(|\Delta S|=1\big) =
\sqrt2\,G_F\,V_{ud}V^{*}_{us}\sum_{i} \widetilde{C}_i\,{\cal O}_i\,,
\label{weak-lagr}
\end{equation}
which can be derived with the help of the Wilson operator product
expansion (OPE) from elementary quark processes, with additional gluon
exchanges.
   In the framework of perturbative QCD the coefficients $\widetilde{C}_i$ 
are to be understood as scale and renormalization scheme dependent functions.

  In (\ref{weak-lagr}), ${\cal O}_i$ are the four-quark operators, defined
either by combinations of products of quark currents ($i=1,2,3,4$, 
non-penguin diagrams) or, in case of gluonic ($i=5,6$) and electro-weak 
($i=7,8$) penguin operators, by products of quark densities.
   The operators ${\cal O}_i$ with $i=1,2,3,5,6$ describe weak transitions
with isospin change $\Delta I=1/2$ while the operator ${\cal O}_4$ corresponds
to $\Delta I=3/2$ transition and operators ${\cal O}_{7,8}$ -- to mixture of 
$\Delta I=1/2$ and $\Delta I=3/2$ amplitudes.

   The general scheme of the meson matrix element calculation by using
of the weak lagrangian (\ref{weak-lagr}) is based on the quark
bosonization approach \cite{bos-weak}. 
   The bosonization procedure establishes the correspondence between 
quark and meson currents (densities) and product of currents (densities).
   Finally, it leads to the effective lagrangian for nonleptonic kaon decays 
in terms of bosonized (meson) currents and densities:
$$
\bar q \gamma_\mu \frac14 (1\mp \gamma^5) \lambda^a q\,\,\, 
\Rightarrow\,\, J^{a\,(mes)}_{L/R \, \mu}\,, 
\quad
\bar q \frac14 (1\mp \gamma^5) \lambda^a q\,\,\,
\Rightarrow\,\, J^{a\,(mes)}_{L/R}\,.
$$

   The meson currents/densities $J^a_{L/R\mu}$ and $J^a_{L/R}$ are obtained 
from the quark determinant by variation over additional external sources
associated with the corresponding quark currents and densities \cite{bos-weak}.
   From the momentum expansion of the quark determinant to $O(p^{2n})$
one can derive the strong lagrangian for mesons ${\cal L}_{eff}$ of
the same order and the corresponding currents and densities $J^a_{L/R\mu}$ 
and $J^a_{L/R}$ to the order $O(p^{2n-1})$ and $O(p^{2n-2})$, respectively.
   Thus, the bosonization approach gives us the correspondence between
power counting for the momentum expansion of the effective chiral
lagrangian of strong meson interactions,
$$
{\cal L}^{(mes)}_s  =
{\cal L}_s^{(p^2)} + {\cal L}_s^{(p^4)} + {\cal L}_s^{(p^6)} + \,.\,.\,.\,,
$$ 
and power counting for the meson currents and densities:
\begin{eqnarray*} 
{\cal L}_s^{(p^n)}\,\,\Rightarrow\,\,J^{(p^{n-1})}_\mu\,\,\mbox{(currents)}\,;
\quad 
{\cal L}_s^{(p^n)}\,\,\Rightarrow\,\,J^{(p^{n-2})}\,\,\mbox{(densities)}\,.
\end{eqnarray*} 

   Some interesting observations on the difference of the momentum behavior
of penguin and non-penguin operators can be drawn from power-counting
arguments.
   The leading contributions to the vector currents and scalar densities are 
of $O(p^1)$ and $O(p^0)$, respectively.
   Since in our approach the non-penguin operators are constructed out of
the products of currents $J^a_{L\mu}$, while the penguin operators
are products of densities $J^a_L$, the lowest-order contributions
of non-penguin and penguin operators are of $O(p^2)$ and $O(p^0)$,
respectively.
   However, due to the well-known cancelation of the contribution of the
gluonic penguin operator ${\cal O}_5$ at the lowest order 
\cite{chivukula}, the leading gluonic penguin as well as non-penguin
contributions start from $O(p^2)$
\footnote{There is no cancellation of the contribution of the
          electromagnetic penguin operator ${\cal O}_8$ at the lowest
          order and the leading contributions start in this case from 
          $O(p^0)$}.
   Consequently, in order to derive the currents which contribute 
to the non-penguin transition operators at the leading order, it is sufficient 
to use the terms of the quark determinant to $O(p^2)$ only.
   At the same time the terms of the quark determinant to $O(p^4)$ have to 
be kept for calculating the penguin contribution at $O(p^2)$, since it
arises from the combination of densities, which are of $O(p^0)$ and 
$O(p^2)$, respectively.
   In this subtle way the difference in momentum behavior is revealed
between matrix elements for these two types of weak transition operators;
it manifests itself more drastically in higher-order lagrangians and
currents.

   This fact makes penguins especially sensitive to higher order effects.
   In particular, the difference in the momentum power counting behavior
between penguin and non-penguin contributions to the isotopic amplitudes
of $K\to 3\pi$ decays, which appears in higher orders of chiral theory,
leads to the dynamical enhancement of the charge asymmetry of the Dalitz-plot
linear slope parameter \cite{CP-enhancement,bos-weak}.

   In our approach the Wilson coefficients $\widetilde{C}_i$ in the effective
weak lagrangian (\ref{weak-lagr}) are treated as the phenomenological 
parameters which should be fixed from the experiment. 
   They are related with the Wilson coefficients $\widetilde{C}^{QCD}_i(\mu)$,
calculated in perturbative QCD \cite{buras3}, via the 
$\widetilde{B}_i$-factors: 
$$
\widetilde{C}_i^{ph}=\widetilde{C}^{QCD}_i(\mu)\widetilde{B}_i(\mu)\,.
$$
   The coefficients $\widetilde{C}^{QCD}_i(\mu)$ contain a small imaginary 
parts and can be presented in the form of a sum of $z$ and $y$ components
\begin{equation}
\widetilde{C}_i^{QCD}(\mu) = \widetilde{C}^{(z)}_i(\mu)
                             +\tau\,\widetilde{C}^{(y)}_i(\mu),\quad
\tau = -\frac{V_{td}V_{ts}^{*}}{V_{ud}V_{us}^{*}}\,.
\label{wilson}
\end{equation}
   Respectively, the amplitudes of nonleptonic kaon decays also contain
$z$- and $y$- components,
\begin{equation}
{\cal A} = {\cal A}^{(z)}+\tau{\cal A}^{(y)}\,,
\label{amplitude}
\end{equation}
which can be expressed in terms of $z$- and $y$-components of Wilson 
coefficients.
    The dominating contributions ${\cal A}^{(i)}$ of the four-quark 
operators ${\cal O}_i$ and $\widetilde{B}_i$-factors may be written as
\begin{eqnarray*}
{\cal A}^{(z,y)}&=&
   \Big[-\widetilde{C}_1^{(z,y)}(\mu)+\widetilde{C}_2^{(z,y)}(\mu)
        +\widetilde{C}_3^{(z,y)}(\mu)\Big]
    \widetilde{B}_1(\mu)\,{\cal A}^{(1)}
\\ &&
  +\widetilde{C}_4^{(z,y)}(\mu)\widetilde{B}_4(\mu)\,{\cal A}^{(4)}
  +\widetilde{C}_5^{(z,y)}(\mu)\widetilde{B}_5(\mu)\,{\cal A}^{(5)}
  +\widetilde{C}_8^{(z,y)}(\mu)\widetilde{B}_8(\mu)\,{\cal A}^{(8)}\,.
\label{components}
\end{eqnarray*}

    The observable effects of direct CP-violation in the nonleptonic kaon 
decays are caused by the $y$-components of their amplitudes (\ref{amplitude}).
    In particular, the ratio $\varepsilon^{'}/\varepsilon$ can be expressed as
$$
\mbox{Re}\bigg( \frac{\varepsilon^{'}}{\varepsilon} \bigg) =
\mbox{Im}\,\lambda_t\,\big(P_0-P_2),\quad
P_I = \frac{\omega}{\sqrt{2}|\varepsilon| |V_{ud}||V_{us}|}\,
      \frac{{\cal A}_I^{(y)}}{{\cal A}_I^{(z)}}\,,
$$
where $\mbox{Im}\,\lambda_t = \mbox{Im}\,V^{*}_{ts}V_{td}
                            = |V_{ub}||V_{cb}| \mbox{sin} \delta$;
$\omega = {\cal A}_2^{(z)}/{\cal A}_0^{(z)}$, and ${\cal A}_I$ are isotopic
amplitudes of $K\to 2\pi$ decay corresponding to the $\pi\pi$ final states 
with the isospin $I=0,2$.

    There are the following theoretical uncertainties which appear both from
short distance (Wilson coefficients) and long distance (effective chiral 
lagrangians and $\widetilde{B}_i$-factors) contributions to the kaon decay
amplitude:
\begin{itemize}
\item Dependence on the parameter $\mbox{Im}\,\tau \sim \mbox{Im}\,\lambda_t$ 
      arising from the imaginary part of Wilson coefficient (\ref{wilson}).

\item Regularization scheme dependence which arises when calculating Wilson
      coefficients beyond the leading order (LO) of QCD in the next-to-leading
      orders: naive dimensional regularization (NDR), t'Hooft-Veltman 
      regularization (HV).

\item Dependence of Wilson coefficients on choice of the renormalization point
      $\mu$ and QCD scale $\Lambda^{(4)}_{\overline{MS}}=(325\pm 110)$ MeV.

\item Dependence of meson matrix elements on the structure constants $L_2$, 
      $L_3$, $L_4$, $L_5$, $L_8$ of the general form of the effective chiral
      lagrangian introduced at $O(p^4)$ by Gasser and Leutwyler \cite{gasser1}.

\item Dependence of meson matrix elements on the structure constants of the 
      effective chiral lagrangian at $O(p^6)$ \cite{p6-our}.

\item Factors $\widetilde{B}_i$ $(i=1,4,5,8)$ for dominating contributions of 
      four-quark operators ${\cal O}_i$ to the meson matrix element
      (\ref{components}).

\item Dependence on choice of the regularization scheme to fix the UV 
      divergences resulting from meson loops.

\end{itemize}

    In the present paper we have combined a new systematic calculation of 
mesonic matrix elements for nonleptonic kaon decays from the effective chiral 
lagrangian approach with Wilson coefficients $\widetilde{C}^{QCD}_i(\mu)$, 
derived by the Munich group \cite{buras3} for $\mu =1$ GeV and $m_t=170$ GeV.
    For the parameter $\mbox{Im}\,\lambda_t$ we have used the result
obtained in \cite{buras5}: $\mbox{Im}\,\lambda_t = (1.33 \pm
0.14)\cdot 10^{-4}$.
    We performed a complete calculation of $K\to 2\pi$ and $K\to 3\pi$
amplitudes at $O(p^6)$ including the tree level, one- and two-loop diagrams.
    For the structure coefficients $L_i$ of the effective chiral lagrangian at 
$O(p^4)$ we used the values fixed in \cite{dafne} from the phenomenological
analysis of low-energy meson processes.
    The structure coefficients of the effective chiral lagrangian at $O(p^6)$ 
have been fixed theoretically from the modulus of the logarithm of the quark 
determinant of the NJL-type model (see \cite{p6-our} for more details).
    The superpropagator regularization has been applied to fix UV divergences
in meson loops.
    The isotopic symmetry breaking ($\pi^0$-$\eta$-$\eta^{'}$ mixing) was
taken into account at the tree level.
    In \cite{k2pi-our} one can find more technical details of the calculation
of $K\to 2\pi$ amplitudes.
  
    In our phenomenological analysis the results on the $K\to 2\pi$ 
amplitudes, including the value for $\varepsilon^{'}/\varepsilon$,
are compared with the experiment to fix the phenomenological 
$\widetilde{B}_i$-factors for the mesonic matrix elements of nonpenguin and 
penguin four-quark operators. 
    As experimental input we used the experimental values of the isotopic 
amplitudes ${\cal A}_{0,2}^{(exp)}$ fixed from the widths of $K\to 2\pi$ 
decays and the world average value 
$\mbox{Re}\,\varepsilon^{'}/\varepsilon = (19.3\pm 3.8)\times 10^{-4}$
which includes both old results of NA31 \cite{NA31-CP} and E731 \cite{E731-CP}
experiments and recent results from KTeV \cite{KTeV-CP} and NA48 
\cite{NA48-CP}.
    The output parameters of the performed $K\to 2\pi$ analysis are the 
factors $\widetilde{B}_1$, $\widetilde{B}_4$ and $\widetilde{B}_5$ for
a fixed value of $\widetilde{B}_8$.

\begin{figure}
\epsfysize=0.8\textheight
\hspace*{15mm}\epsfbox{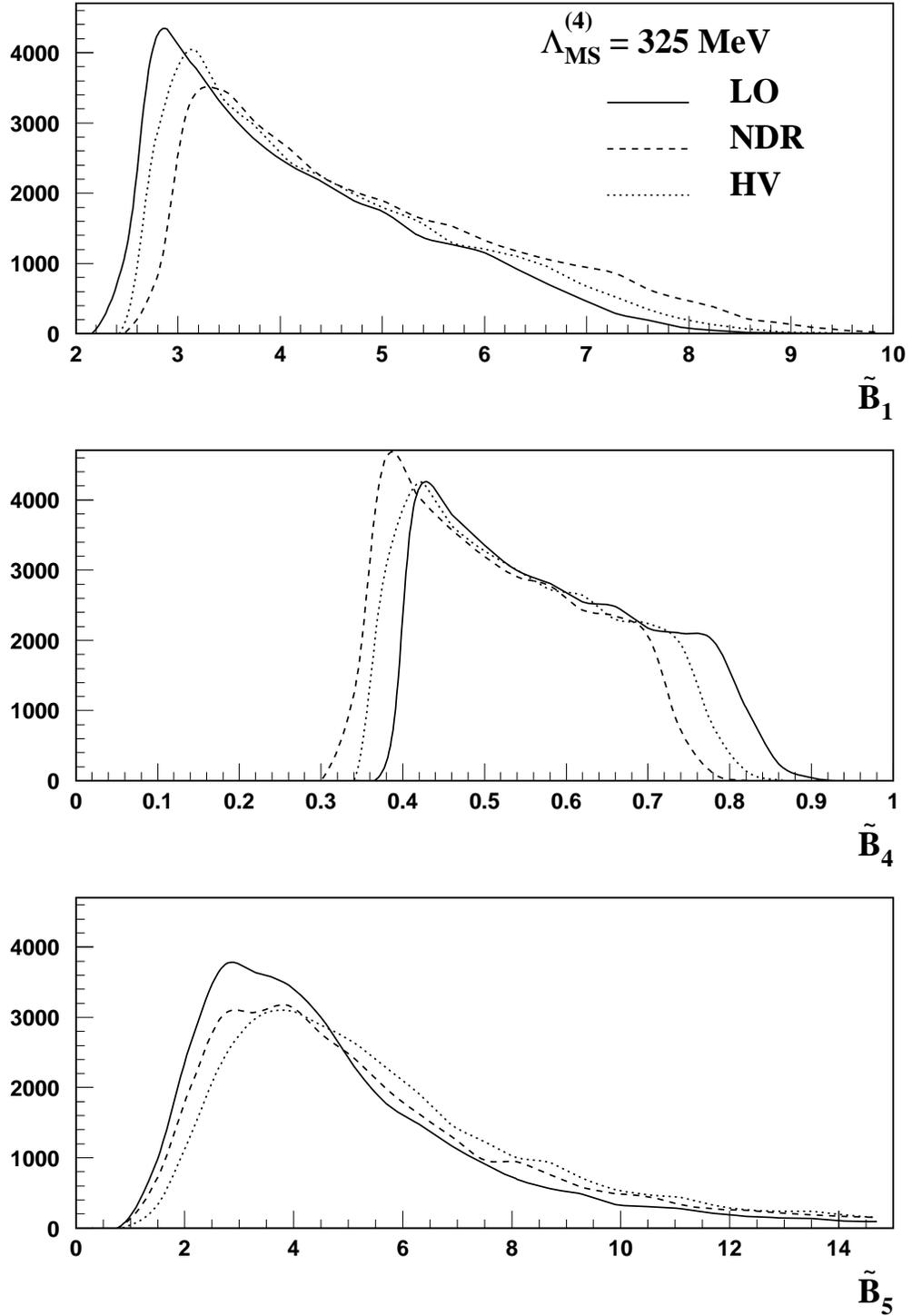}
\caption{%
Probability density distributions for factors
$\widetilde{B}_1$, $\widetilde{B}_4$ and $\widetilde{B}_5$ 
with $\widetilde{B}_8=1$.}
\label{fig1}
\end{figure}
\begin{figure}
\epsfysize=0.2\textheight
\hspace*{15mm}\epsfbox{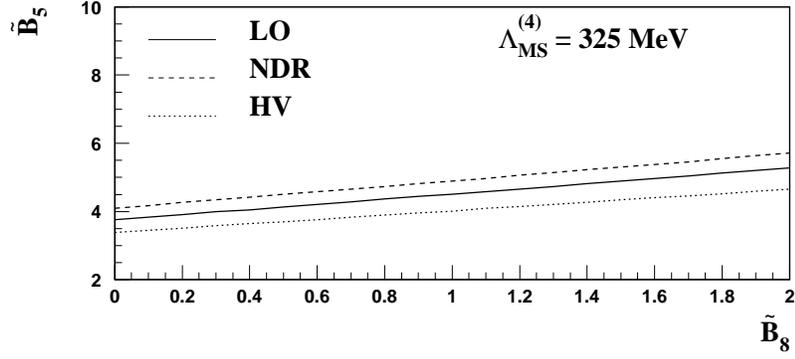}
\caption{Correlations between parameters $\widetilde{B}_5$ and $\widetilde{B}_8=1$.}
\label{fig2}
\end{figure}
\begin{figure}
\vspace*{-4mm}
\epsfysize=0.2\textheight
\hspace*{20mm}\epsfbox{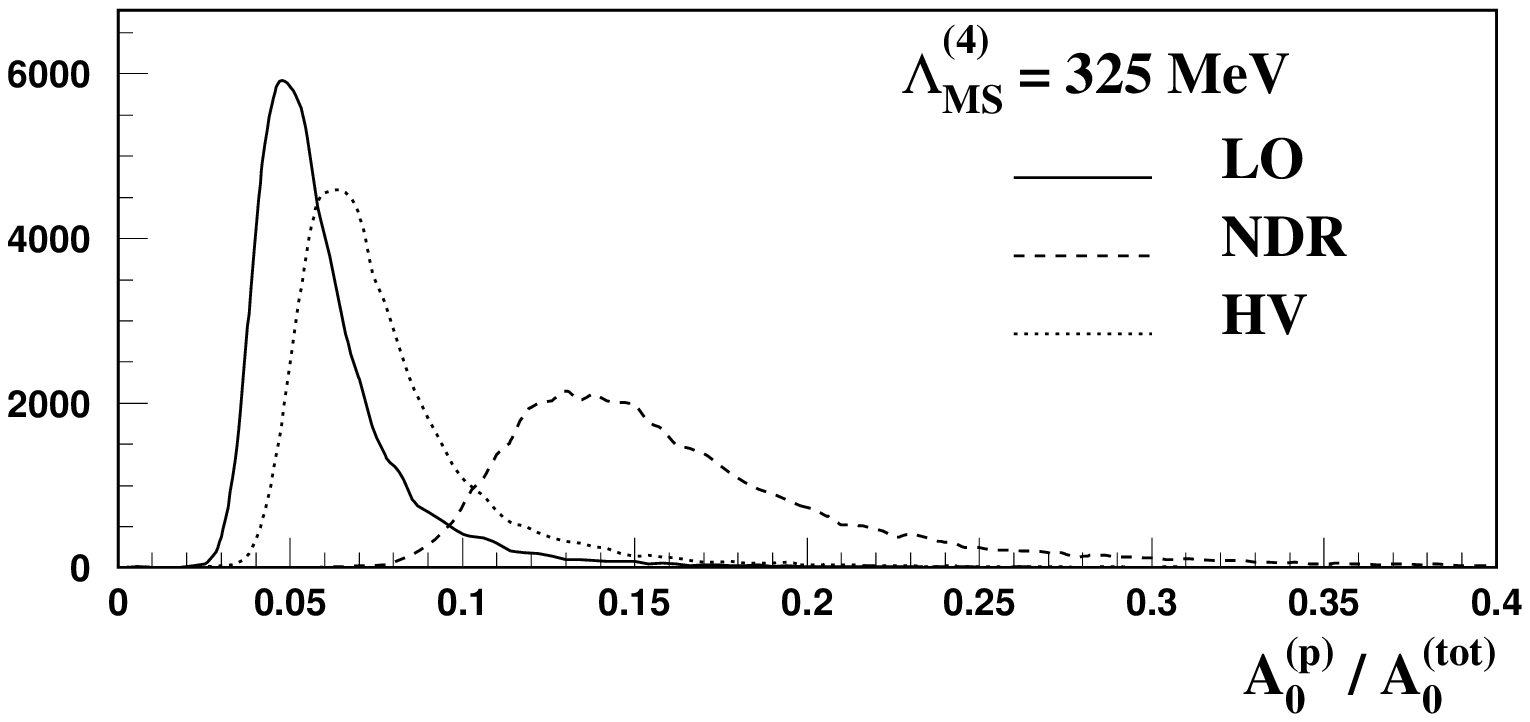}
\caption{%
Probability density distributions for the relative
contribution of penguin operators to the $\Delta I = 1/2$ amplitude.}
\label{fig3}
\end{figure}

    The dependence of $\widetilde{B}_i$-factors on  different theoretical 
uncertainties and experimental errors of various input parameters is 
investigated by applying the ``Gaussian'' method. 
    Using  Wilson coefficients derived in \cite{buras3} in various
regularization schemes (LO, NDR, HV) for different values of the QCD scale 
$\Lambda^{(4)}_{\overline{MS}}$ ,we calculated the probability density 
distributions for $\widetilde{B}_i$-factors obtained by using Gaussian 
distribution for all input parameters with their errors.
    As an example, the probability densities for the parameters 
$\widetilde{B}_1$, $\widetilde{B}_4$, $\widetilde{B}_5$ calculated with
$\widetilde{B}_8=1$ and $\Lambda^{(4)}_{\overline{MS}}=325$ MeV are shown 
in Figure~\ref{fig1}.
    We have shown the necessity for a rather large gluonic penguin 
contribution to describe the recently confirmed large experimental
$\varepsilon^{'}$ value (the factor $\widetilde{B}_5$ is found well above 1). 
    Figure~\ref{fig2} shows the correlations between $\widetilde{B}_5$ and 
$\widetilde{B}_8$ calculated for central values of all input parameters.
    From this figure one can see that even for $\widetilde{B}_8=0$ values of
$\widetilde{B}_5 > 2$ are necessary to explain the large value
of $\varepsilon^{'}/\varepsilon$.
    It should be emphasized, that for even larger values of $\widetilde{B}_5$,
the contributions of nonpenguin operators to the $\Delta I = 1/2$ amplitude 
are still dominating (see Figure~\ref{fig3}).
    The large $\widetilde{B}_1$ and $\widetilde{B}_5$ values may be a hint 
that the long-distance contributions, especially to $\Delta I = 1/2$ 
amplitudes, are still not completely understood.
    An analogous conclusion has been drawn in \cite{buras5}, where
possible effects from physics beyond the Standard Model are also discussed.

    Finally, we present our predictions for the CP-asymmetry of linear 
slope parameters in the $K^{\pm}\to 3\pi$ Dalitz plot.
    These predictions are based on a new calculation of $K\to 3\pi$
amplitudes at $O(p^6)$ within the same effective lagrangian approach.
    The obtained  $K\to 3\pi$ amplitudes include the same theoretical
uncertainties as in case of the $K\to 2\pi$ analysis.  
    The values of $\widetilde{B}_1$, $\widetilde{B}_4$, $\widetilde{B}_5$
fixed from the $K\to 2\pi$ analysis are used as phenomenological input
to the $K\to 3\pi$ estimates which have been performed in a self-consistent
way by the Gaussian method.

\begin{figure}
\epsfysize=0.4\textheight
\hspace*{22mm}\epsfbox{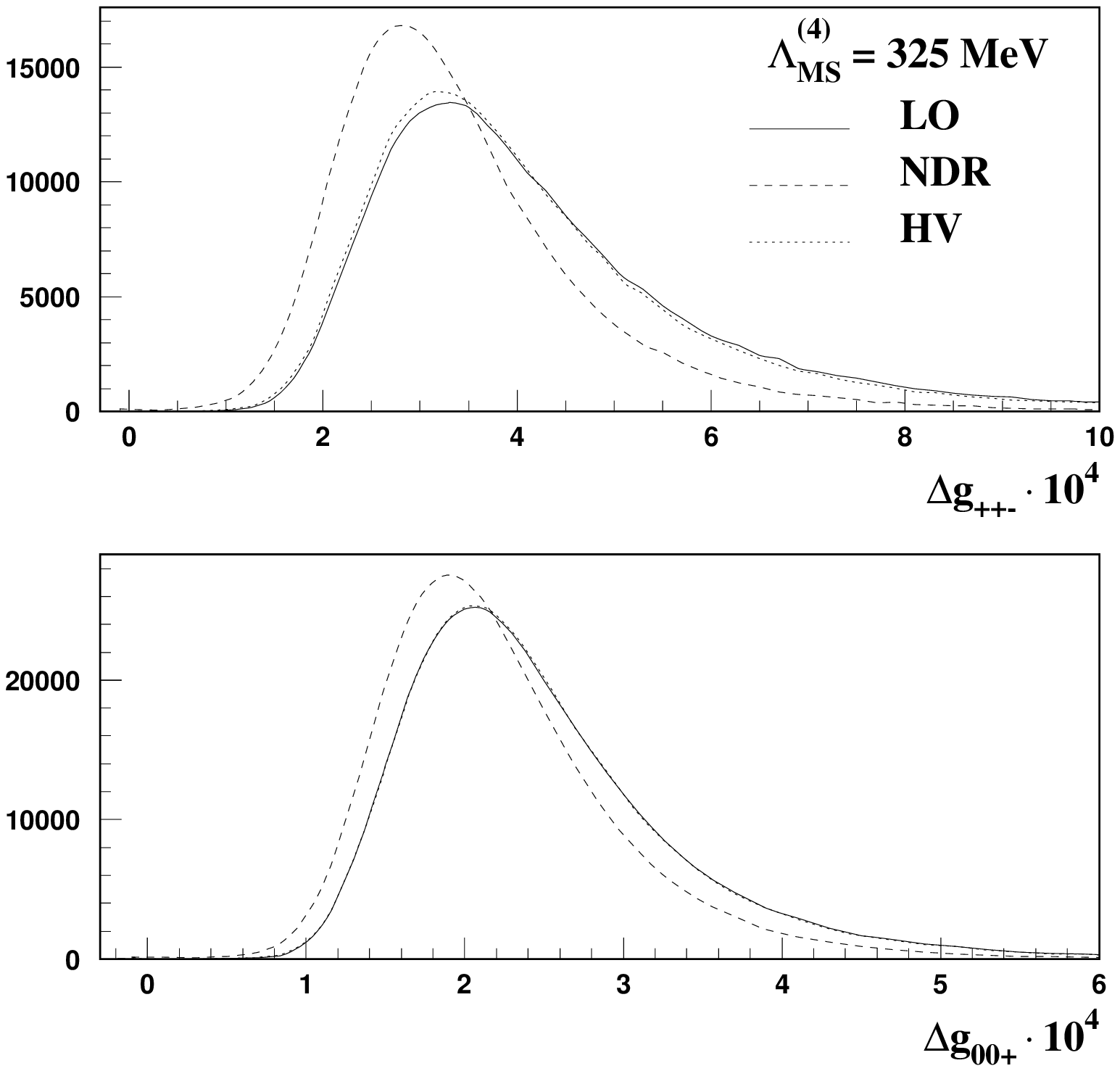}
\caption{%
Probability density distributions for the CP-asymmetry of 
linear slope parameters of 
$K^{\pm}\to\pi^{\pm}\pi^{\pm}\pi^{\mp}$ and
$K^{\pm}\to\pi^{0}\pi^{0}\pi^{\pm}$ decays.}
\label{fig4}
\end{figure}

    The linear slope parameter $g$ of the Dalitz plot for $K\to 3\pi$ decays 
is defined through the expansion of the decay probability over the kinematic
variables
$$
 |T(K\to 3\pi)|^2\,\,\propto\,\, 1 + gY +\,.\,.\,. 
$$
where $Y$ is a Dalitz variable:
$$
  Y = \frac{s_3-s_0}{m^2_\pi}\,,\quad
  s_3 = (p_K -p_{\pi_3})^2\,,\quad
  s_0 =  \frac{m^2_K}{3} + m^2_\pi\,,
$$
and the notation $\pi_3$ belongs to the non-equivalent pion in the decays
$K^{\pm}\to \pi^{\pm}\pi^{\pm}\pi^{\mp}$ and 
$K^{\pm}\to \pi^{0}\pi^{0}\pi^{\pm}$.
    Direct CP-violation leads to a charge asymmetry of the linear slope 
parameter,
$$
\Delta g(K^\pm \to 3\pi) = \frac{g(K^{+}\to 3\pi)-g(K^{-}\to 3\pi)}
                                {g(K^{+}\to 3\pi)+g(K^{-}\to 3\pi)}\,.
$$

    In Figure~\ref{fig4} we show the probability density distributions for 
$K^{\pm}\to\pi^{\pm}\pi^{\pm}\pi^{\mp}$ and 
$K^{\pm}\to\pi^{0}\pi^{0}\pi^{\pm}$ decays calculated with
$\widetilde{B}_8=1$ and $\Lambda^{(4)}_{\overline{MS}}=325$ MeV.
    Upper and low bounds for $\Delta g_{++-}$ and $\Delta g_{00+}$ for 
different values of $\Lambda^{(4)}_{\overline{MS}}$ in LO, NDR and HV
regularization schemes ($\widetilde{B}_8 = 1$) obtained by the Gaussian method 
are shown in table 1.
    The limits without brackets correspond to the confidence level of 68\% 
while the limits in brackets -- to the confidence level of 95\%.  
    Summarizing these results, we have obtained the following upper and lower 
bounds for the charge symmetries of the linear slope parameter:
$$
      2.1\, <\,\Delta g_{++-}\cdot 10^4\,<\, \,6.2\,,\quad
      1.4\, <\,\Delta g_{00+}\cdot 10^4\,<\, \,3.5\,\,\,\, 
      \mbox{with\,\,CL=68\%};
$$
$$ 
      1.4\, <\,\Delta g_{++-}\cdot 10^4\,< \,10.4\,,\quad
      1.0\, <\,\Delta g_{00+}\cdot 10^4\,< \,5.1\,\,\,\,
      \mbox{with\,\,CL=95\%}.
$$

\vspace{3mm}
\begin{center}
Table 1. Upper and low bounds for $\Delta g_{++-}$ and $\Delta g_{00+}$ 
         (in units $10^{-4}$).

\vspace{3mm}
{\small
\begin{tabular}{|c|c|*{2}{c}|*2{c}|*2{c}|} \hline \hline
$~\Delta g$& $\Lambda ^{(4)}_{\overline{MS}}$ & 
\multicolumn{2}{|c|}{LO}&\multicolumn{2}{|c|}{NDR}&\multicolumn{2}{|c|}{HV}\\
\cline{3-8}
$~~~~~$&(MeV)&  min &  max &  min &  max &  min &  max   \\
\hline
       & 215 &  2.5 & 5.9  &  2.1 & 4.7  &  2.6 & 6.0    \\
       &     &( 1.8 & 10.1)&( 1.5 & 7.3 )&( 1.9 & 10.2 ) \\ \cline{2-8}
$\Delta g_{++-}$ 
       & 325 &  2.6 & 6.0  &  2.2 & 4.7  &  2.6 & 5.8    \\
       &     &( 1.9 & 10.2)&( 1.5 & 7.2 )&( 1.8 & 9.7  ) \\ \cline{2-8}
       & 435 &  2.7 & 6.2  &  2.2 & 4.7  &  2.6 & 5.8    \\
       &     &( 1.9 & 10.4)&( 1.4 & 7.0) &( 1.8 & 9.5  ) \\ \hline \hline

       & 215 &  1.5 & 3.3  &  1.4 & 2.9  &  1.6 & 3.4    \\
       &     &( 1.1 & 4.8 )&( 1.0 & 4.2 )&( 1.2 & 5.1 )  \\ \cline{2-8}
$\Delta g_{00+}$ 
       & 325 &  1.6 & 3.4  &  1.4 & 3.0  &  1.6 & 3.3    \\ 
       &     &( 1.2 & 5.0 )&( 1.0 & 4.3 )&( 1.2 & 4.9 )  \\ \cline{2-8}
       & 435 &  1.7 & 3.5  &  1.5 & 3.1  &  1.7 & 3.4    \\
       &     &( 1.2 & 5.1 )&( 1.0 & 4.3 )&( 1.2 & 5.0 )  \\ \hline \hline
\end{tabular}
}
\end{center}

    With some experimental updates and theoretical refinements our new 
estimates confirm the dynamical enhancement mechanism for the charge asymmetry
$\Delta g$ by higher order contributions in the effective chiral lagrangian 
approach which was first observed in \cite{CP-enhancement}.
    The predicted slope parameter asymmetry, although small, may be in reach
of current high statistics experiments \cite{app}.
    Already with lower statistics, new measurements of quadratic slope 
parameters of $K\to 3\pi$ decays, including the neutral channels, would lead 
to the improved theoretical understanding of the nonperturbative part of 
nonleptonic kaon decay dynamics.


\end{document}